# Second order perturbed Heisenberg Hamiltonian of $Fe_3O_4$ ultra-thin films

P. Samarasekara

Department of Physics, University of Peradeniya, Peradeniya, Sri Lanka

**Abstract**

Due to the wide range of applications, theoretical models of $Fe_3O_4$ films are found to be important. Ultra thin $Fe_3O_4$ films with ferrite structure have been theoretically investigated using second order perturbed modified Heisenberg Hamiltonian. Matrices for ultra thin films with two and three spin layers are presented in this manuscript. Total magnetic energy was expressed in terms of spin exchange interaction, magnetic dipole interaction, second order magnetic anisotropy and stress induced magnetic anisotropy. Magnetic properties were observed for films with two spin layers and variant second order magnetic anisotropy. For the film with three spin layers, second order anisotropy constant was fixed to avoid tedious derivations. Magnetic easy axis rotates toward the in plane direction as the number of spin layers is increased from two to three because the stress induced anisotropy energy dominates at higher number of spin layers. According to some other experimental data, the magnetic easy axis of thin films rotates toward the in plane direction as the thickness is increased. For ferrite film with two spin layers, magnetic easy and hard directions can be observed at 0.75 and 1.2 radians, respectively, when the ratio of stress induced anisotropy to the long range dipole interaction strength is 3.9. For ferrite film with three spin layers, magnetic easy and hard directions can be observed at 2.4 and 2.3 radians, respectively, when the ratio of stress induced anisotropy to the long range dipole interaction strength is 4.2.

## 1. Introduction:

Magnetite ($Fe_3O_4$) and maghemite ($\gamma Fe_2O_3$) are the most popular natural oxides. $Fe_3O_4$ finds potential applications in magnetic storage, industrial catalysts, water purification and drug delivery. $Fe_3O_4$ is a ferrite with inverse spinel structure. Spinel structure with tetrahedral and octahedral sites can be found in detail in some previous publications [1, 2, 3, 4, 5]. Five of $Fe^{3+}$ ions occupy tetrahedral sites. Other five $Fe^{3+}$ ions and four $Fe^{2+}$ ions occupy octahedral sites. Because magnetic moments of $Fe^{3+}$ in tetrahedral and octahedral sites cancel each other, the net magnetic moment of $Fe_3O_4$ is completely due to the magnetic moments of four $Fe^{2+}$ ions. Therefore, the theoretical net magnetic moment of $Fe_3O_4$ is 4 Bohr magnetons. However, experimental value of net magnetic moment is approximately 4.1 Bohr magnetons. The spinel structure of this ferrite is represented by $Fe^{3+}(Fe^{2+}Fe^{3+})O_4$. The magnetic moments of $Fe^{2+}$ and $Fe^{3+}$ are 4 $\mu_B$ and 5 $\mu_B$, respectively.

Cation distribution of ferrite like compounds has been found using Rietveld method [1, 2]. Surface spin waves in CsCl type ferrimagnet with a (001) surface has been studied by combining Green function theory with the transfer matrix method [6]. Anisotropy of ultrathin ferromagnetic films and the spin reorientation transition have been investigated using Heisenberg Hamiltonian with few terms [7]. In addition, the surface magnetism of ferrimagnet thin films has been studied using Heisenberg method [8]. The surface spin wave spectra of both the simple cubic and body centered ferrimagnets have been theoretically studied using Heisenberg Hamiltonian [9]. The cation distribution and oxidation state of Mn-Fe spinel nanoparticles have been systematically studied at various temperatures by using neutron diffraction and electron energy loss spectroscopy [4]. The crystal structure of spinel type compounds has been found using single crystal X-ray diffraction data [3]. The lattice parameter, anion parameter and the cation inversion parameter of spinel structures have been presented [5]. Surface spin waves on the (001) free surface of semi-infinite two lattice ferrimagnets on the Heisenberg model with nearest neighbor exchange interactions has been investigated [10].

Ferromagnetic ultra-thin and thick films have been investigated using second order perturbed Heisenberg Hamiltonian by us [11]. Previously ferromagnetic ultra thin and thick films have been studied using third order perturbed Heisenberg Hamiltonian [12, 13]. Furthermore, ferrite ultra-thin and thick films have been investigated using second order perturbed Heisenberg Hamiltonian by us [14, 15]. Ferrite ultra-thin and thick films have been investigated using third order perturbed Heisenberg Hamiltonian [16, 17]. In this manuscript, the spinel structure of $Fe_3O_4$ was used to find the magnetic properties of $Fe_3O_4$ ultra thin films.

## 2. Model:

Classical Heisenberg Hamiltonian of a thin film can be written as following.

$$H = -J \sum_{m,n} \vec{S}_m \cdot \vec{S}_n + \omega \sum_{m \neq n} \left( \frac{\vec{S}_m \cdot \vec{S}_n}{r_{mn}^3} - \frac{3(\vec{S}_m \cdot \vec{r}_{mn})(\vec{r}_{mn} \cdot \vec{S}_n)}{r_{mn}^5} \right) - \sum_m D_{\lambda_m}^{(2)} (S_m^z)^2$$

$$- \sum_m K_s Sin 2\theta_m \qquad (1)$$

Here J, ω, θ, $D_m^{(2)}$, $K_s$, m, n and N are spin exchange interaction, strength of long range dipole interaction, azimuthal angle of spin, second order anisotropy constant, stress induced anisotropy constant, spin plane indices and total number of spin layers in film, respectively. When the stress applies normal to the film plane, the angle between m[th] spin and the stress is $θ_m$.

The cubic cell was divided into 8 spin layers with alternative $Fe^{2+}$ and $Fe^{3+}$ spins layers (Sickafus et al. 1999). The spins of $Fe^{2+}$ and $Fe^{3+}$ will be taken as 1 and p, respectively. While the spins in one layer point in one direction, spins in adjacent layers point in opposite directions. A thin film with (001) spinel cubic cell orientation will be considered. The length of one side of unit cell will be taken as "a". Within the cell the spins orient in one direction due to the super exchange interaction between spins (or magnetic moments). Therefore, the results proven for oriented case in one of our early report [15] will be used for following equations. But the angle θ will vary from $\theta_m$ to $\theta_{m+1}$ at the interface between two cells.

For a thin film with thickness Na,

Spin exchange interaction energy=$E_{exchange}$= $N(-10J+72Jp-22Jp^2)+8Jp\sum_{m=1}^{N-1}\cos(\theta_{m+1}-\theta_m)$

Dipole interaction energy=$E_{dipole}$

$$E_{dipole} = -48.415\omega\sum_{m=1}^{N}(1+3\cos2\theta_m) + 20.41\omega p\sum_{m=1}^{N-1}[\cos(\theta_{m+1}-\theta_m)+3\cos(\theta_{m+1}+\theta_m)]$$

Here the first and second term in each above equation represent the variation of energy within the cell and the interface of the cell, respectively.

Total energy

E= $N(-10J+72Jp-22Jp^2)+8Jp\sum_{m=1}^{N-1}\cos(\theta_{m+1}-\theta_m)$

$-48.415\omega\sum_{m=1}^{N}(1+3\cos2\theta_m) + 20.41\omega p\sum_{m=1}^{N-1}[\cos(\theta_{m+1}-\theta_m)+3\cos(\theta_{m+1}+\theta_m)]$

$$-\sum_{m=1}^{N}[D_m^{(2)}\cos^2\theta_m]-4(1-p)\sum_{m=1}^{N}[K_s\sin 2\theta_m] \qquad (2)$$

Here the anisotropy energy term and the last term have been explained in our previous report for oriented spinel ferrite [15]. If the angle is given by $\theta_m=\theta+\varepsilon_m$ with perturbation $\varepsilon_m$, after taking the terms up to second order perturbation of $\varepsilon$ only,

The total energy can be given as $E(\theta)=E_0+E(\varepsilon)+E(\varepsilon^2)$

Here

$E_0= -10JN+72pNJ-22Jp^2N+8Jp(N-1)-48.415\omega N-145.245\omega N\cos(2\theta)$

$$+20.41\omega p[(N-1)+3(N-1)\cos(2\theta)]-\cos^2\theta\sum_{m=1}^{N}D_m^{(2)}-4(1-p)NK_s\sin 2\theta \qquad (3)$$

$$E(\varepsilon)=290.5\omega\sin(2\theta)\sum_{m=1}^{N}\varepsilon_m-61.23\omega p\sin(2\theta)\sum_{m=1}^{N-1}(\varepsilon_m+\varepsilon_n)$$

$$+\sin 2\theta\sum_{m=1}^{N}D_m^{(2)}\varepsilon_m-8(1-p)K_s\cos 2\theta\sum_{m=1}^{N}\varepsilon_m \qquad (4)$$

$$E(\varepsilon^2)=-4Jp\sum_{m=1}^{N-1}(\varepsilon_n-\varepsilon_m)^2+290.5\omega\cos(2\theta)\sum_{m=1}^{N}\varepsilon_m^2-10.2\omega p\sum_{m=1}^{N-1}(\varepsilon_n-\varepsilon_m)^2$$

$$-30.6\omega p\cos(2\theta)\sum_{m=1}^{N-1}(\varepsilon_n+\varepsilon_m)^2-(\sin^2\theta-\cos^2\theta)\sum_{m=1}^{N}D_m^{(2)}\varepsilon_m^2$$

$$+8(1-p)[K_s\sin 2\theta\sum_{m=1}^{N}\varepsilon_m^2] \qquad (5)$$

The sin and cosine terms in equation number 2 have been expanded to obtain above equations.

Here n=m+1.

Under the constraint $\sum_{m=1}^{N}\varepsilon_m = 0$, first and last terms of equation 4 are zero.

Therefore, $E(\varepsilon) = \vec{\alpha}.\vec{\varepsilon}$

Here $\vec{\alpha}(\varepsilon) = \vec{B}(\theta)\sin 2\theta$ are the terms of matrices with

$$B_\lambda(\theta) = -122.46\omega p + D_\lambda^{(2)} \quad (6)$$

Also $E(\varepsilon^2) = \frac{1}{2}\vec{\varepsilon}.C.\vec{\varepsilon}$, and matrix C is assumed to be symmetric ($C_{mn}=C_{nm}$).

Here the elements of matrix C can be given as following,

$C_{m, m+1}$=8Jp+20.4ωp-61.2pωcos(2θ)

For m=1 and N,

$C_{mm}$= -8Jp-20.4ωp-61.2pωcos(2θ)+581ωcos(2θ) $- 2(\sin^2\theta - \cos^2\theta) D_m^{(2)}$

$\quad +16(1-p)[K_s \sin(2\theta)] \quad (7)$

For m=2, 3, ----, N-1

$C_{mm}$= -16Jp-40.8ωp-122.4pωcos(2θ)+581ωcos(2θ) $- 2(\sin^2\theta - \cos^2\theta) D_m^{(2)}$

$\quad +16(1-p)[K_s \sin(2\theta)]$

Otherwise, $C_{mn}=0$

Therefore, the total energy can be given as

$$E(\theta)=E_0+ \vec{\alpha}.\vec{\varepsilon} +\frac{1}{2}\vec{\varepsilon}.C.\vec{\varepsilon} =E_0-\frac{1}{2}\vec{\alpha}.C^+.\vec{\alpha} \tag{8}$$

Here $C^+$ is the pseudo-inverse given by

$$C.C^+ =1-\frac{E}{N}. \tag{9}$$

Here E is the matrix with all elements $E_{mn}=1$.

## 3. Results and discussion:

A film with two spin layers (N=2) will be considered first. If the anisotropy constants vary within the film, then $C_{12}=C_{21}$ and $C_{11} \neq C_{22}$.

Then $C^+_{11} = -C^+_{12} = \dfrac{C_{22}+C_{21}}{2(C_{11}C_{22}-C_{21}^2)}$ and $C^+_{21} = -C^+_{22} = \dfrac{C_{21}+C_{11}}{2(C_{21}^2-C_{11}C_{22})}$.

Hence, $\vec{\alpha}.C^+.\vec{\alpha} = (\alpha_1-\alpha_2)(C^+_{21}\alpha_2 - C^+_{12}\alpha_1)$

Then

$C_{11}$= -8Jp-20.4ωp-61.2pωcos(2θ)+581ωcos(2θ) + 2(cos 2θ) $D_1^{(2)}$ +16(1-p)$K_s$sin(2θ)

$C_{22}$= -8Jp-20.4ωp-61.2pωcos(2θ)+581ωcos(2θ) + 2(cos 2θ) $D_2^{(2)}$ +16(1-p)$K_s$sin(2θ)

$C_{12} = 8Jp + 20.4\omega p - 61.2p\omega\cos(2\theta)$

$\alpha_1 = [-122.46\omega p + D_1^{(2)}]\sin(2\theta)$

$\alpha_2 = [-122.46\omega p + D_2^{(2)}]\sin(2\theta)$

$E(\theta) = E_0 - \dfrac{(\alpha_1 - \alpha_2)(C^+_{21}\alpha_2 - C^+_{12}\alpha_1)}{2}$

$E_0 = -20J + 144pJ - 44Jp^2 + 8Jp - 96.83\omega - 290.5\omega\cos(2\theta) + 20.41\omega p[1 + 3\cos(2\theta)]$

$-\cos^2\theta[D_1^{(2)} + D_2^{(2)}] - 4(1-p)NK_s\sin(2\theta)$

Here $D_1^{(2)}$ and $D_2^{(2)}$ were taken as the anisotropy constants of first and second spin layers of the $Fe_3O_4$ film, respectively.

Because the magnetic moments of $Fe^{2+}$ and $Fe^{3+}$ are 4 $\mu_B$ and 5 $\mu_B$, respectively,

p=5/4=1.25.

Then

$C_{11} = -10J - 25.5\omega + 504.5\omega\cos(2\theta) + 2(\cos 2\theta)D_1^{(2)} - 4K_s\sin(2\theta)$

$C_{22} = -10J - 25.5\omega + 504.5\omega\cos(2\theta) + 2(\cos 2\theta)D_2^{(2)} - 4K_s\sin(2\theta)$

$C_{12} = 10J + 25.5\omega - 76.5\omega\cos(2\theta)$

$E_0 = 101.25J - 71.32\omega - 214\omega\cos(2\theta) - \cos^2\theta[D_1^{(2)} + D_2^{(2)}] + 2K_s\sin(2\theta)$

Figure 1 shows the 3-D plot of $\frac{E(\theta)}{\omega}$ versus θ and $\frac{K_s}{\omega}$, for $\frac{J}{\omega} = \frac{D_1^{(2)}}{\omega} = 10, \frac{D_2^{(2)}}{\omega} = 5$. Minima of this 3-D plot can be observed at $\frac{K_s}{\omega}$ =2.9, 3.9, 7, ------ etc. Maxima of this 3-D plot can be observed at 1.9, 4.0, 4.9, ------- etc. According to this graph, film can be easily oriented in some particular directions by applying a stress. The total energy of this ferrite ultra thin film is much smaller than the total energy of thick ferromagnetic films implying that total energy increases with the number of spin layers [12]. However, this graph is similar to the 3-D graph of $\frac{E(\theta)}{\omega}$ versus θ and $\frac{K_s}{\omega}$ obtained for nickel ferrite [15]. Figure 2 shows the graph of $\frac{E(\theta)}{\omega}$ versus angle for $\frac{K_s}{\omega}$ =3.9. One minimum and a consecutive maximum of this graph can be observed at 0.75 and 1.2 radians, respectively. Energy minima and maxima correspond to magnetic easy and hard directions, respectively. Changing the value of $\frac{K_s}{\omega}$ didn't change this graph of energy versus angle considerably. In addition, several less spaced peaks can be observed in this case compared to thick ferromagnetic films [12]. However, the spikes observed in energy versus angle graph of thin ferromagnetic films with two layers don't appear in this graph [13].

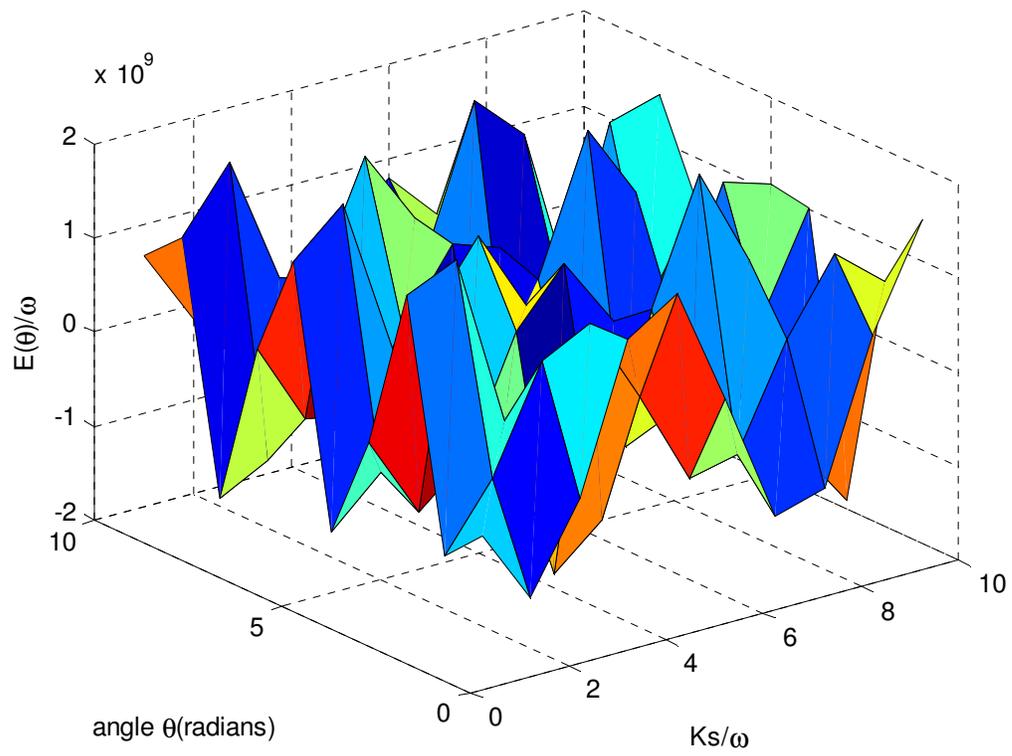

Figure 1: 3-D plot of energy versus angle and stress induced anisotropy for N=2.

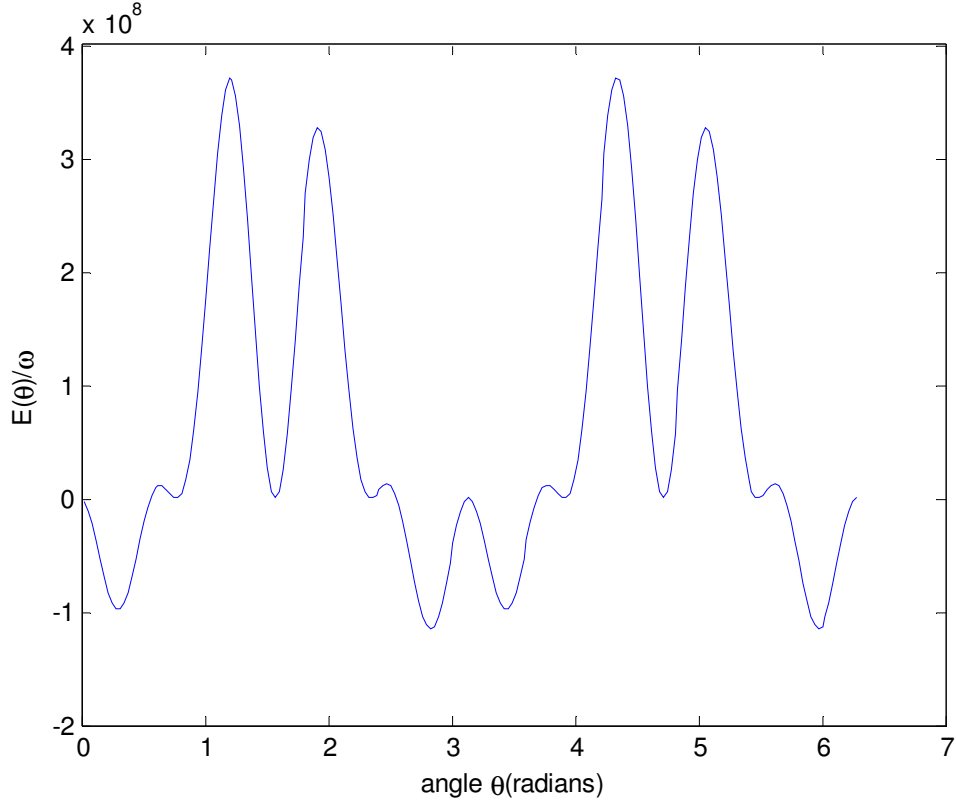

Figure 2: Graph of energy versus angle for $\frac{K_s}{\omega}=3.9$ and N=2.

For N=3, the each $C^+_{nm}$ element found using equation 9 is consist of more than 20 terms. To avoid this problem, matrix elements were found using $C.C^+=1$. Then $C^+_{mn}$ is given by $C^+_{mn} = \frac{cofactor C_{nm}}{\det C}$. Under this condition, $\vec{E}.\vec{\alpha} = 0$, and the average value of first order perturbation is zero. The second order anisotropy constant is assumed to be an invariant for the convenience.

Then $C_{11}=C_{33}$, $C_{12}=C_{21}=C_{23}=C_{32}$, $C_{13}=C_{31}=0$, $\alpha_1=\alpha_2=\alpha_3$.

$C_{11}=C_{33}=$ -10J-25.5$\omega$+504.5$\omega$cos(2$\theta$) + 2(cos 2$\theta$) $D_m^{(2)}$ -4$K_s$sin(2$\theta$)

$C_{22}= -20J-51\omega+428\omega\cos(2\theta) + 2(\cos 2\theta) D_m^{(2)} -4K_s\sin(2\theta)$

Therefore, $C^+_{11} = \dfrac{C_{11}C_{22} - C_{32}^2}{C_{11}^2 C_{22} - 2C_{32}^2 C_{11}} = C^+_{33}$, $C^+_{13} = \dfrac{C_{32}^2}{C_{11}^2 C_{22} - 2C_{32}^2 C_{11}} = C^+_{31}$

$C^+_{12} = \dfrac{-C_{32}C_{11}}{C_{11}^2 C_{22} - 2C_{32}^2 C_{11}} = C^+_{21} = C^+_{23} = C^+_{32}$, $C^+_{22} = \dfrac{C_{11}^2}{C_{11}^2 C_{22} - 2C_{32}^2 C_{11}}$

The total energy can be found using the following equation.

$E(\theta)=E_0-0.5[C^+_{11}(2\alpha_1^2)+C^+_{32}(4\alpha_1^2)+C^+_{31}(2\alpha_1^2)+\alpha_1^2 C^+_{22}]$

Here $E_0= 156.875J-94.22\omega-282.66\omega\cos(2\theta) - \cos^2\theta[D_1^{(2)} + D_2^{(2)} + D_3^{(2)}] +3K_s\sin(2\theta)$

Figure 3 shows the 3-D plot of $\dfrac{E(\theta)}{\omega}$ versus $\theta$ and $\dfrac{K_s}{\omega}$, for $\dfrac{J}{\omega} = \dfrac{D_m^{(2)}}{\omega} = 10$. Maxima of this graph can be observed at $\dfrac{K_s}{\omega}$=5.2, 8.2, ---- etc. Minima of the graphs can be observed at $\dfrac{K_s}{\omega}$=4.2, 6.2, 8.2, -----etc. The total energy of ferrite thin films with three layers obtained using third order perturbation is comparable to the total energy in this case [16]. The total energy of thick ferrite films obtained using third order perturbed Heisenberg Hamiltonian is higher than that of this ultra thin ferrite film [17]. Figure 4 shows the graph of $\dfrac{E(\theta)}{\omega}$ versus angle for $\dfrac{K_s}{\omega}$=4.2. Nearest maxima and minima of this graph can be observed at 2.3 and 2.4 radians, respectively. In this case, the magnetic hard and easy directions are 2.3 and 2.4, respectively. The position of minima and maxima of these graphs don't change considerably with $\frac{K_s}{\omega}$. The shape of energy versus angle graph of ferrite thin films with three layers obtained using third order perturbation is different from the same graph in this case [16]. Some spikes observed in energy versus angle graph of ferromagnetic films with five layers don't appear in this graph. According to our previous experimental data of ferrite thin films, the coercivity and the magnetic anisotropy depend on the stress of the film [18].

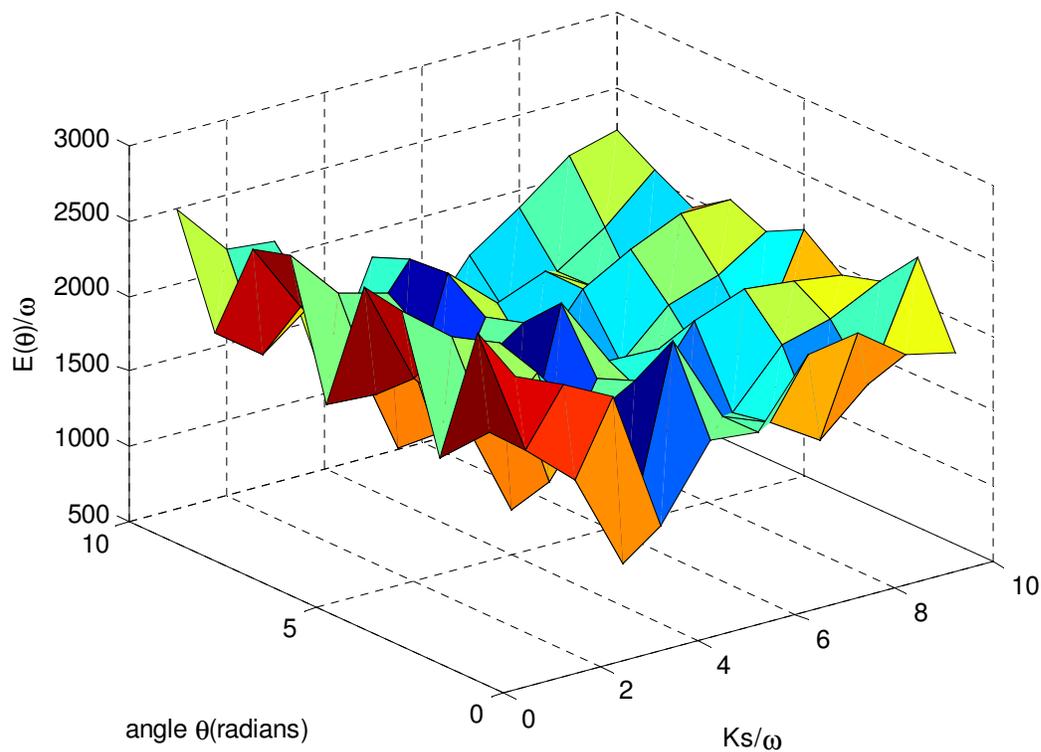

Figure 3: 3-D plot of energy versus angle and stress induced anisotropy for N=3.

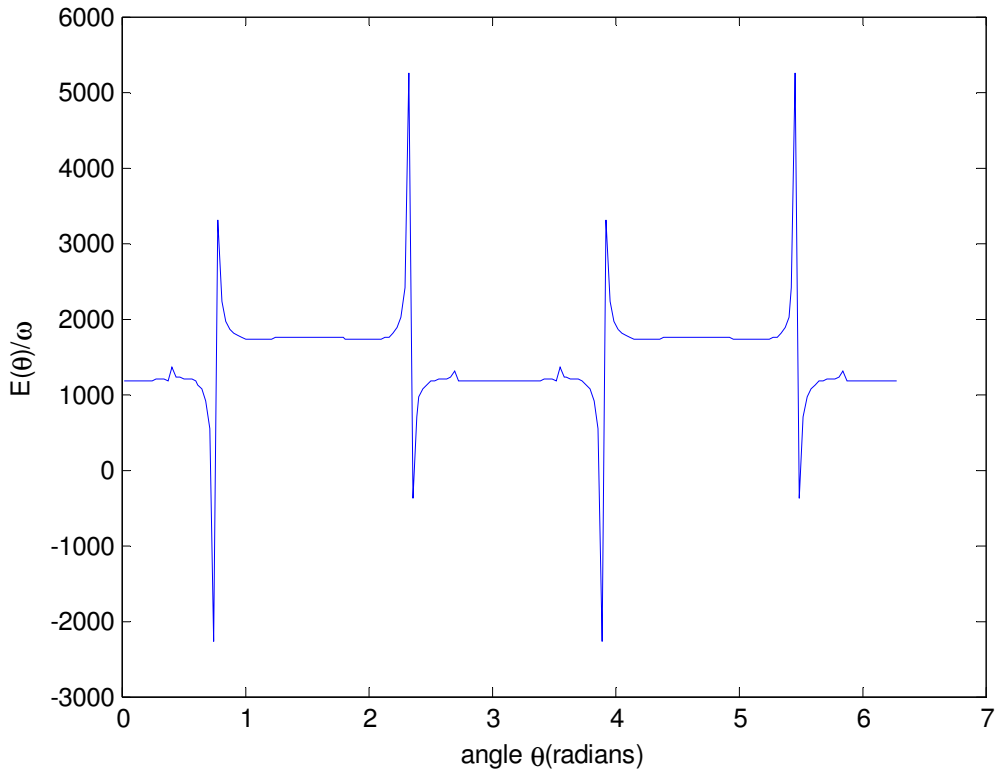

Figure 4: Graph of energy versus angle for $\frac{K_s}{\omega}$ =4.2 and N=3.

According to figures 2 and 4, the first magnetic easy direction can be observed at 0.25 and 0.75 radians for two and three spin layers, respectively. Therefore, the magnetic easy axis rotates toward the in plane direction as the number of layers is increased. According to the equation of total energy, only the $E_0$ term depends on the number of layers (N). Only term with N is the stress induced anisotropy. Because 1-p is negative for most of the spinel ferrites, the energy of stress induced anisotropy increases with the number of layers. As the thickness is increased, stress induced anisotropy dominates spin exchange interaction, second order magnetic anisotropy and magnetic dipole interaction. The domination of stress induced anisotropy is the possible reason for the rotation of easy direction with the number of layers. The same phenomenon was observed for thick ferromagnetic films using second order and third order perturbed Heisenberg Hamiltonian by us. According to some experimental data, the easy axis of magnetic thin films rotates toward the in plane direction, as the number if layers is increased [19, 20]. In addition, the magnetic easy direction of thin films depends on the deposition temperature. The variation of easy axis with deposition temperature can be explained using spin reorientation coupled with Heisenberg Hamiltonian [11, 21, 22].

## 4. Conclusion:

Total magnetic energy, magnetic easy direction and magnetic hard direction were determined for films with two and three spin layers by plotting 3-D graphs of energy versus stress induced anisotropy and angle, and graphs of energy versus angle. Minimum of energy of 3-D plot can be observed at several values of stress induced anisotropy. For two spin layers, consecutive minimum and maximum can be observed at 0.75 and 1.2 radians, respectively for $\frac{K_s}{\omega}$=3.9. For three spin layers, consecutive minimum and maximum can be observed at 2.4 and 2.3 radians, respectively for $\frac{K_s}{\omega}$=4.2. According to the first minima of 2-D plots of two and three layers, the magnetic easy axis gradually rotates toward the in plane direction of the film as the number of spin layers is increased. The possible reason for the rotation of magnetic easy axis is the domination of stress induced magnetic anisotropy. This theoretical data qualitatively agree with some experimental data of magnetic thin films.